\RequirePackage{fix-cm}
\documentclass[twocolumn]{svjour3}          
\smartqed  

\usepackage{lineno}
\modulolinenumbers[5]
\usepackage{graphicx}
\usepackage{amsmath, amssymb}
\usepackage{mathtools}
\usepackage{algorithm}
\usepackage{algpseudocode}
\usepackage{multirow}
\usepackage{array}
\usepackage[hidelinks]{hyperref}
\usepackage{subfigure}
\usepackage{listings}
\usepackage{color}
 
\definecolor{codegreen}{rgb}{0,0.6,0}
\definecolor{codegray}{rgb}{0.5,0.5,0.5}
\definecolor{codepurple}{rgb}{0.58,0,0.82}
\definecolor{backcolour}{rgb}{0.95,0.95,0.92}
 
\lstdefinestyle{mystyle}{
    commentstyle=\color{codegreen},
    keywordstyle=\color{magenta},
    numberstyle=\tiny\color{codegray},
    stringstyle=\color{codepurple},
    basicstyle=\footnotesize,
    breakatwhitespace=false,         
    breaklines=true,                 
    captionpos=b,                    
    keepspaces=true,                 
    numbers=left,                    
    numbersep=5pt,                  
    showspaces=false,                
    showstringspaces=false,
    showtabs=false,                  
    tabsize=2
}
 
\newcommand{\ndlib}{{\scshape NDlib}}
\newcommand{\dynetx}{{\scshape DyNetX}}
\begin{document}

\title{\ndlib: a Python Library to Model and Analyze Diffusion Processes Over Complex Networks\footnote{This paper is an extension version of the DSAA'2017 Application Track paper titled ``\ndlib: Studying Network Diffusion Dynamics"}
}


\author{Giulio Rossetti        \and
        Letizia Milli \and\\
        Salvatore Rinzivillo \and
        Alina S\^irbu \and \\
        Dino Pedreschi \and
        Fosca Giannotti
}


\institute{Letizia Milli, Alina S\^irbu and Dino Pedreschi \at
              University of Pisa,\\
              Largo Bruno Pontecorvo, 2 Pisa, Italy\\ 
              \email{name.surname@di.unipi.it}           
           \and
           Giulio Rossetti, Letizia Milli, Salvatore Rinzivillo and Fosca Giannotti \at
             KDD Lab. ISTI-CNR,\\ 
via G. Moruzzi, 1 Pisa, Italy\\
   \email{name.surname@isti.cnr.it}
}

\date{Received: date / Accepted: date}

\maketitle

\begin{abstract}
Nowadays the analysis of dynamics of and on networks represents a hot topic in the Social Network Analysis playground.
To support students, teachers, developers and researchers in this work we introduce a novel framework, namely \ndlib, an environment designed to describe diffusion simulations.
\ndlib\ is designed to be a multi-level ecosystem that can be fruitfully used by different user segments.
For this reason, upon \ndlib, we designed a simulation server that allows remote execution of experiments as well as an online visualization tool that abstracts its programmatic interface and makes available the simulation platform to non-technicians. 
\end{abstract}

\section{Introduction}
In the last decades Social Network Analysis, henceforth SNA, has received increasing attention from several, heterogeneous fields of research.
Such popularity was certainly due to the flexibility offered by graph theory: a powerful tool that allows reducing countless phenomena to a common analytical framework whose basic bricks are nodes and their relations.
Indeed, social relationships, trading, transportation and communication infrastructures, even the brain can be modeled as networks and, as such, analyzed.
Undoubtedly, such pervasiveness has produced an amplification in the visibility of network analysis studies thus making this complex and interesting field one of the most widespread among higher education centers, universities and academies.
Given the exponential diffusion reached by SNA, several tools were developed to make it approachable to the wider audience possible.
SNA programming libraries are nowadays available to computer scientists, physicists as well as mathematicians; moreover, graphical tools were developed for social scientists, biologists as well as for educational purposes.

Although being a very active field of research per se, SNA is often used as a tool to analyze complex phenomena such as the unfolding of human interactions, the spreading of epidemic and diffusion of opinions, ideas, innovations.
Even for such peculiar applications, we have witnessed during the last years the appearance of dedicated tools and libraries: however, the plethora of resources available often discourage the final users making hard and time-consuming the identification of the right tool for the specific task and level of expertise.

To cope with such issue, in this work we introduce a novel framework designed to model, simulate and study  diffusive phenomena over complex networks.\\
\ndlib\ represents a multi-level solution targeted to epidemic spreading simulations. 

The final aim of this paper is to introduce \ndlib, describing its main features, and comparing it with other tools already available to SNA practitioners.
One of the main contribution of the proposed tool regards the heterogeneous audience it is intended to serve.
Unlike its competitors, \ndlib\ is designed to provide access to diffusion models to both technicians (e.g., programmers or researchers coming from quantitative disciplines) and non-technicians (e.g., students, social science researchers, analysts).
To fulfill such goal it offers, aside from the standard programmatic interface, a web based visual one able to abstract from the low level description and definition of diffusion simulation.
Moreover, \ndlib\ comes with an experiment server platform that allows to easily set up controlled exploratories and to decouple experiment definition and execution.

\ndlib\ is intended to overcome some limitations identified in existing libraries (e.g., providing support for dynamic network topologies) while reducing the overall usage complexity. 

The paper is organized as follows. 
In Section \ref{sec:ndif} we briefly introduce some concepts needed to frame the context \ndlib\ is designed to analyze.
In such section are briefly discussed both dynamics \emph{on} and \emph{of} networks: there, a simple classification of the diffusive process as well as and two classic dynamic network model definition are provided.
Moving from such definitions, in Section \ref{sec:ndlibenv} we introduce \ndlib: there we describe how the library is designed, its rationale and main characteristics.
Moreover, still in Section \ref{sec:ndlibenv}, we introduce \ndlib-REST and \ndlib-Viz: the former being a service designed to offer remote simulation facilities, the latter a web-based visual platform that abstracts from the coding complexity and allows the end user to setup, run and analyze diffusion experiments without writing any line of code.  
Once make clear the key features of \ndlib\, in Section \ref{sec:tools} we identify the competitors of our library: there, for some of them, we propose a two level comparison that takes into account both \emph{qualitative} and \emph{quantitative} analysis.
Indeed, we do not aim to propose an extensive review of existing libraries and tools: our results are conversely intended to allow potential users to synthetically compare some of the most used existing resources.
In Section \ref{sec:conclusion} we conclude the paper, underlining the advantages of \ndlib\ w.r.t. its competitors and providing insights on the future evolution of our framework.
Finally, an appendix to this paper briefly describes the models made available by \ndlib\ v3.0.0.

\section{Dynamics On and Of Networks}
\label{sec:ndif}
Network dynamics is, nowadays, one of the emerging themes within complex network analysis.
Since its beginning, complex network analysis has been approached through the definition of very specific, task-oriented, mining problems. 
The almost complete absence of the time dimension in such problem definitions comes from historical reasons that can be identified in (i) the graph theory ancestry of the field, and in (ii) the reduced number of dynamic data sources available at the time the area of complex networks analysis emerged.

During recent years such scenario radically changed, prevalently due to the explosion of human-generated traces collected via socio-technical platforms: wide repositories of time-aware data that can be easily modeled as dynamic networks.
Indeed, Graphs have often been used to model and study dynamic phenomena: to better describe these realities, in which both relationships among agents as well as their status change through time, several works in the last few years have started to lay the foundations of temporal network analysis.

In this section we introduce the basic concepts behind network diffusion, \ref{subsec:netd}, and dynamic network analysis, \ref{subsec:dyna} that are leveraged by \ndlib, thus proposing a contextualization for the models and data structures it defines and uses.

\subsection{Network Diffusion}
\label{subsec:netd}
The analysis of diffusive phenomena that unfold on top of complex networks is a task attracting growing interests from multiple fields of research.
The models proposed in the literature to address such  extensively studied problem can be broadly separated into two, strictly related, classes: \emph{Epidemics and Spreading} and \emph{Opinion Dynamics}.
In the following we will briefly introduce such categories, describing their common aspects as well as differences.
Indeed, it is clear that the two categories have enough in common  to be implemented under a common framework, which is why we decided to introduce both of them within \ndlib.

\subsubsection{Epidemics and Spreading}
When we talk about epidemics, we think about contagious diseases caused by biological pathogens, like influenza, measles, chickenpox and  sexually transmitted viruses that spread from person to person. 
However, other phenomena can be linked to the concept of epidemic: think about the spread of computer virus \cite{Szor04} where the agent is a malware that can transmit a copy of itself from computer to computer, or the spread of mobile phone virus \cite{Havlin2009,IJInfSec_WGMB-Virus}, or the diffusion of knowledge, innovations, products in an online social network \cite{Burt:1987}, the so-called ``social contagion", where people are making decision to adopt a new idea or innovation.
Several elements determine the patterns by which epidemics spread through groups of people: the properties carried by the pathogen  (its contagiousness, the length of its infectious period and its severity), the structure of the network as well as the mobility patterns of the people involved.
Although often treated as similar processes, diffusion of information and epidemic spreading can be easily distinguished by a single feature: the degree of \emph{activeness} of the subjects they affect \cite{MRPG17}.

Indeed, the spreading process of a virus does not require an \emph{active} participation of the people that catch it (i.e., even though some behaviors acts as contagion facilitators -- scarce hygiene, moist and crowded environment -- we can assume that no one chooses to get the flu on purpose); conversely, we can argue that the diffusion of an idea, an innovation, or a trend strictly depend not only on the social pressure but also by individual choices.
For a list of epidemic and spreading models refer to the Appendix.

\subsubsection{Opinion Dynamics}
A different field related to modeling social behavior is that of opinion dynamics. 
Recent years have witnessed the introduction of a wide range of models that attempt to explain how opinions form in a population \cite{sirbu2017opinion}, taking into account various social theories (e.g., bounded confidence \cite{Deffuant2000} or social impact \cite{Sznajd-Weron2001}). 
These models have a lot in common with those seen in epidemics and spreading. 
In general, individuals are modeled as agents with a state and connected by a social network.  
The social links can be represented by a complete graph (mean field models) or by more realistic complex networks. 
Node state is typically represented by variables that can be discrete (similar to the case of spreading) but also continuous, representing, for instance, a probability of choosing one option or another \cite{sirbu2013opinion}.  
The state of individuals changes in time, based on a set of update rules, mainly through interaction with the neighbors. 
While in many spreading and epidemics models this change is irreversible (susceptible to infected), in opinion dynamics the state can oscillate freely between the possible values, simulating thus how opinions change in reality.  
A different important aspect in opinion dynamics is external information, which can be interpreted as the effect of mass media. 
In general, external information is represented as a static individual with whom all others can interact, again present also in spreading models.
For a list of opinion dynamics models refer to the Appendix.

\subsection{Time-Evolving Networks}
\label{subsec:dyna}

A significant number of social systems can, potentially, be modeled as temporal networks: cellular processes, social communications, economic transactions are only a few examples of contexts that posses both network and temporal aspects that make them attractive for temporal network modeling.

The emergence for dynamic networks analysis theoretical grounds has been highlighted in the book ``Temporal Networks" \cite{Holme2013} where the curators, Holme and Saramaki, proposed an ensemble of works covering different dynamic network analysis methodologies. 
As a first step, several works have tried to transpose known problems on static networks to temporal networks: Temporal motifs mining \cite{kovanen2011temporal}, Diffusion \cite{lee2012exploiting,pennacchioli2013three}, Link prediction \cite{tabourier2016predicting,rossetti2016supervised}, community discovery \cite{rossetti2016tiles,rossetti2017community,rossetti2017rdyn}, are only a few examples.

In a dynamic context, all network entities, nodes and edges (also called \emph{interactions}, due to their dynamic nature), can vary as time goes by. 
In a social scenario -- such as the one described by telephone call logs -- dynamic networks flexibility naturally models the volatility of human interactions whose valence/duration tend to be overestimated when their analysis is made through classical graph theory tools.

Moreover, to support the definition of such revised analytical framework, several formalisms have been proposed to represent evolving network topologies without loss of information: Temporal Networks \cite{Holme2013}, Time-Varying Graphs \cite{casteigts2012time}, Temporal/Interaction Networks \cite{rossetti2016supervised}, and Link Streams \cite{VLM15}, to name the most famous. 
Henceforth, we use the term \textit{Dynamic Network} to encompass all those formalisms.


\subsubsection{Dynamic Network Models}
\label{subsec:dyth}
As previously discussed, the complex systems that we refer to as \emph{dynamic networks} can be modeled following different paradigms.
Among them two are the ones that more often are used in time-aware analytical processes: \emph{Temporal Networks} and \emph{Snapshot Graphs}.
\\ \ \\
\noindent{\bf Temporal Networks.}
A Temporal Network describes a dynamic structure in which both nodes and edges may appear and disappear as time goes by. 
It can be more formally defined as \cite{rossetti2017community}:
\begin{definition}[Dynamic Network]
A Dynamic Network is a graph $DG=(V,E,T)$ where: $V$ is a set of triplets of the form $(v,t_s,t_e)$, with $v$ a vertex of the graph and $t_s, t_e \in T$ are respectively the birth and death timestamps of the corresponding vertex (with $t_s \leq t_e$); $E$ is a set of quadruplets $(u,v,t_s,t_e)$, with $u, v \in V$ are vertices of the graph and $t_s, t_e \in T$ are respectively the birth and death timestamps of the corresponding edge (with $t_s \leq t_e$).
\end{definition}
A node or edge can appear several times if it has several periods of existence.
Depending on the interaction semantics, we can deal with undirected Temporal Networks (TN) or with Directed Temporal Networks (DTN).
Modeling a network as a stream of interactions enables online analysis as well as to track and describe at each moment the exact system status. 
No temporal aggregation acts as a proxy for the dynamic process the analyst wants to study: all the (micro) topological perturbations appear instantaneously. 
Obviously, this model is suitable for particular contexts (i.e., phone call records, online communications, \dots) where characteristic stabilization windows do not exist (or are not easy to identify). 

A major advantage of this model is its completeness concerning phenomena description.
On the other hand, this rich temporal-data modeling does not straightforwardly permit to use methodologies and/or measures defined for static graphs.
\\ \ \\
\noindent{\bf Snapshot Graphs.}
Often, network history is partitioned into a series of snapshots, corresponding either to the state of the network at a time $t$ (relation network) or the aggregation of observed interactions during a period \cite{rossetti2017community}.

\begin{definition}[Snapshot Graph]
A Snapshot Graph $\mathcal{SG}_{\tau}$ is defined by an ordered set $\langle G_1,G_2 \dots G_t \rangle$ where each snapshot $G_i=(V_i,E_i)$ is univocally identified by the sets of nodes $V_i$ and edges $E_i$.
\end{definition}

Network snapshots can be effectively used, for instance, to model a phenomenon that generates network perturbations (almost) at regular intervals. 
Context-dependent temporal windows are used, in this scenario, to partition the network history into consecutive snapshots, time-bounded observations describing a precise, static discretization of the network life. 

The snapshot-based analysis is frequently adopted for networks without a natural temporal discretization, due to the balance they propose between model complexity and expressivity. 
They allow to apply static networks tools on evolving networks: algorithms, as well as network measures and definitions, can be independently applied to each temporal partition without the need of novel analytical tools that explicitly manage temporal information.
An issue of this model is related to the identification of the optimal window size to use to generate the set of snapshots, a choice that can deeply affect the outcome of the subsequent analysis.

\section{\ndlib: Network Diffusion Library}
\label{sec:ndlibenv}
Since the analysis of diffusive phenomena represents a hot topic for some communities having different backgrounds, we designed our framework so that it can be fruitfully used by the widest user segment possible.
To do so, we organized it in three incremental modules: the \ndlib\ core library (written in Python), a remote REST-ful experiment server accessible through API calls and, finally, a web oriented visual interface.

In this section we will describe and discuss the major characteristics of our library (as implemented in v3.0.0), paying attention to underline the rationale behind the implementation choices made and their repercussions on the framework usability.

\subsection{Library Rationale}
At the core of our tool there is \ndlib, whose name stands for \emph{``(N)etwork (D)iffusion Library"}, a Python package built upon the network facilities offered by NetworkX\footnote{NetworkX: \url{https://goo.gl/PHXdnL}}.
The library, available for Python 2.7.x and 3.x, is currently hosted on GitHub\footnote{\ndlib\ GitHub: \url{https://goo.gl/zC7p7b}}, on pypi\footnote{\ndlib\ pypi: \url{https://goo.gl/gc96xW}} and has its online documentation on ReadTheDocs\footnote{\ndlib\ docs: \url{https://goo.gl/VLWtrn}}.
Moreover, \ndlib\ is also made available through the SoBigData.eu catalogue\footnote{SoBigData: \url{http://www.sobigdata.eu}}.
A complete list of the diffusion models implemented in \ndlib\ v3.0.0 -- 11 from the epidemics, 5 from the opinion dynamics categories and 3 designed for dynamic networks --, along with their short descriptions, is reported in Appendix.

\ndlib\ models diffusive phenomena as discrete-time agent-based processes. 
Given a network $G=(V,E)$, a diffusion model $M$, and the actual state of its nodes, $S_i$, the request of a diffusion iteration will return a novel nodes' state $S_{i+1}$ obtained by applying the evolution rules specified by $M$ to all the nodes in $V$.

To do so, the library model a generic diffusion process as composed of three elements: (i) the graph topology on which the process takes place; (ii) the diffusion model; (iii) the model configuration and initial infection state.
The first component, the graph topology, is borrowed by the available entities exposed by the NetworkX library. 
The implementation of all \ndlib\ models is agnostic w.r.t. the directedness of the graph, thus allowing the user to use both undirected as well as directed networks in his simulations.
The second component, the model, is designed so to expose to the final user a minimal and coherent interface to select  diffusion processes: all the models provided extend a generic template that takes care of handling model specific meta-data and to expose step-by-step simulation facilities.
Finally, the third component, the simulation configuration interface, allows the user to fully specify three different classes of information:
\begin{itemize}
	\item[(i)] model specific parameters (e.g., the $\beta$ and $\gamma$ parameter for the {\em SIR} model \cite{wkermack27}); 
	\item[(ii)] nodes' and edges' attributes (e.g., node/edge-wise thresholds for the {\em Threshold} \cite{granovetter1978threshold} and {\em Independent Cascade} \cite{Kempe} like models);
	\item[(iii)] the initial state of the epidemic process (provided as the percentage of nodes belonging to each of the model available statuses, or as a planted initial configuration of node statuses). 
\end{itemize}
The configuration component plays a fundamental role in the library logic: it acts as the focus of experiment description, thus making the simulation definition and invocation coherent across all models.

The following code shows an example of experiment definition, configuration and execution.
\vspace{0.15cm}
\begin{lstlisting}[language=Python]
import networkx as nx
import ndlib.models.ModelConfig as mc
import ndlib.models.epidemics.SIRModel as sir

# Network topology
g = nx.erdos_renyi_graph(1000, 0.1)

# Model selection
model = sir.SIRModel(g)

# Model Configuration
cfg = mc.Configuration()
cfg.add_model_parameter('beta', 0.001)
cfg.add_model_parameter('gamma', 0.01)
cfg.add_model_parameter("percentage_infected", 0.05)
model.set_initial_status(cfg)

# Simulation execution
iterations = model.iteration_bunch(200)
\end{lstlisting}
\vspace{0.15cm}
In lines 1-3 are imported all the required modules; in line 6 an Erd\"os Renyi graph \texttt{g} is built using a NetworkX generator; in line 9 the {\em SIR} \texttt{model} is \emph{attached} to the graph \texttt{g}; in lines 12-16 the \texttt{model} initial status is configured; finally, line 19 shows how 200 iterations of the simulation can be obtained by the invocation of the \texttt{model.iteration\_bunch(bunch\_size=n)} method (where $n=200$ identifies the number of desired iterations).
An alternative to the iteration bunch simulation request is offered by the step-by-step \\ \texttt{model.iteration()} method, a call that returns as output a single simulation iteration.
The status returned by both methods is incremental: each iteration describes the configurations of those nodes that changed their status from the previous model iteration.  

\subsection{Visualization Facilities}
To allow the final user to easily analyze simulation evolution,  \ndlib\ exposes a set of visual facilities. 
In \ndlib\ v3.0.0 two different visual context are integrated: Bokeh\footnote{Bokeh: \url{https://goo.gl/VtjuKS}}, that generate javascript enriched visualization for the web, and Matplotlib\footnote{Matplotlib: \url{https://goo.gl/EY96HV}}, whose plots are print oriented. 

\begin{figure*}[t]
  \centering 
\includegraphics[width=.95\columnwidth]{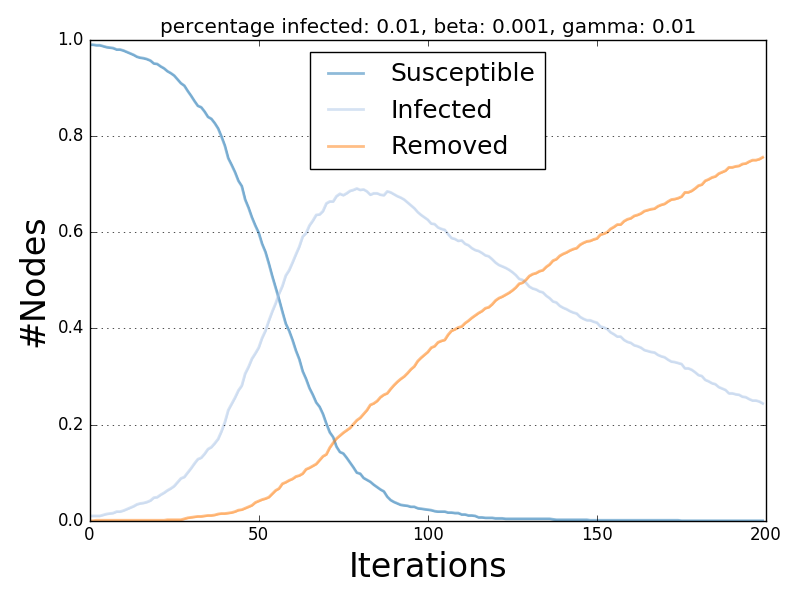}\qquad
\includegraphics[width=.95\columnwidth]{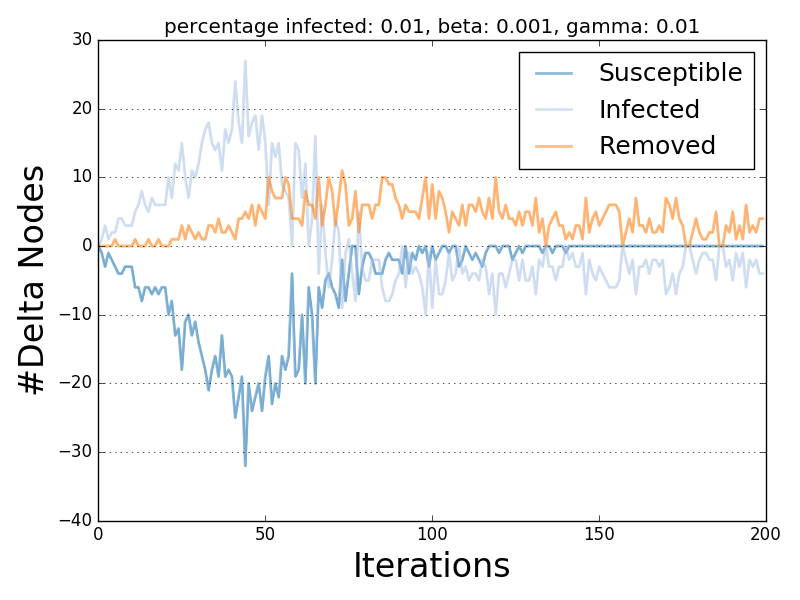}\\
\includegraphics[width=.95\columnwidth]{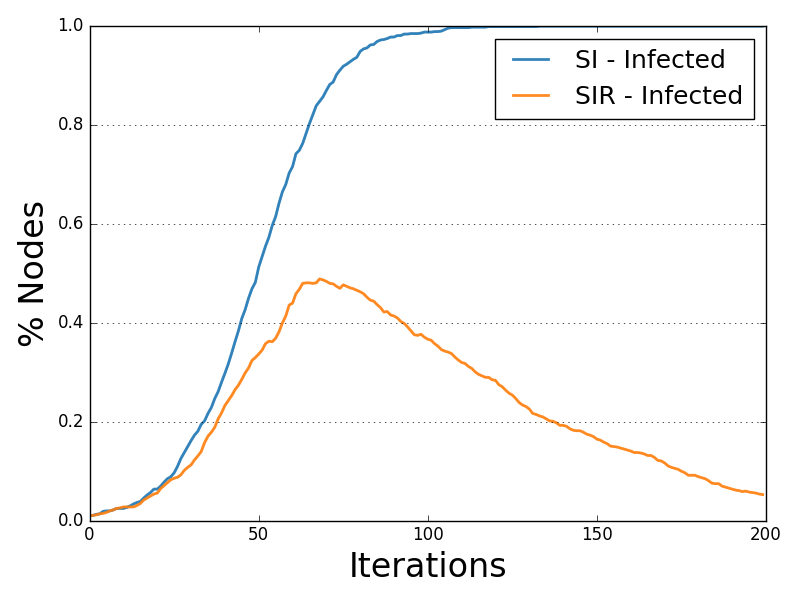}\qquad
\includegraphics[width=.95\columnwidth]{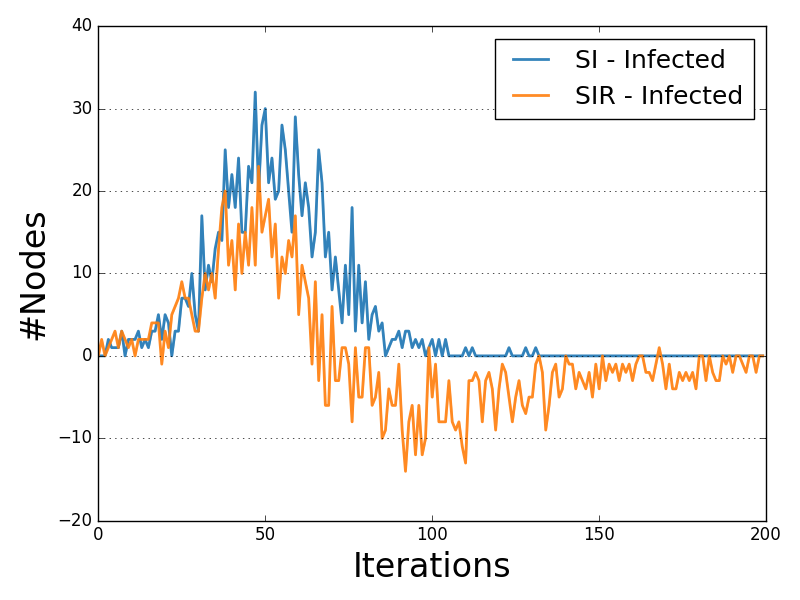}
  \caption{{\bf Visual Analysis.} \ndlib\ \texttt{DiffusionTrend} and \texttt{DiffusionPrevalence} SIR plots. \texttt{DiffusionTrendComparison} and \texttt{DiffusionPrevalenceComparison} for the ``Infected" class of two distinct models.}
  \label{img:diff_trend}
\end{figure*}

Matplotlib plotting facilities are exposed through the visualization package \texttt{ndlib.viz.mpl}.
Among the visual facilities offered by such package, the Diffusion Trend plots describe the evolution of a diffusion process as time goes by from the classes distribution point of view (Figure \ref{img:diff_trend}, first row, left). 
The second basilar plot provided by the Matplotlib interface is the Diffusion Prevalence one (Figure \ref{img:diff_trend}, first row, right) that captures, for each iteration, the variation of the nodes of each class.
To generate such line plots \ndlib\ leverages the meta-data of the applied diffusion model and the resulting simulation iterations: labels, headings, parameter values are retrieved by the configured model object thus making the plot template agnostic w.r.t. the specific diffusion model used during the simulation.

Moreover, to compare the executions of different models \texttt{ndlib.viz.mpl} implements two additional endpoints: Diffusion Trend Comparison and  Diffusion Prevalence Comparison.
Assuming to have executed two or more models on the same network, each of such classes can be used to produce a comparative plot that generalizes the two previously introduced.
Moreover, comparison plots allow selecting the statuses to compare across the models. 
Figures \ref{img:diff_trend}, second row, illustrate  examples of comparison plots built over the results of SI and SIR models for the \emph{Infected} node class.

Bokeh visualizations are exposed through the\\  \texttt{ndlib.viz.bokeh} package and implement the same kind of plotting facilities, following a same programmatic template. 
The only difference between the two visual contexts, apart from the specific target of the produced graphics, lies in the comparison plots. 
Differently, from Matplotlib, Bokeh does not allow to plot multiple trends on the same plot. 
As an alternative solution, such visualization context exposes a multiplot facility with grid auto-layout.

All the visualization offered by \ndlib\ extends a common abstract template. 
Such implementative choice -- borrowed from what already done for the diffusion models definition -- allows the library users to easily extend its plotting functionality by following a predefined schema.

\subsection{Model Parallel Executions}
The evolution of diffusive processes is strictly tied to three factors: (i) the underlying network topology; (ii) the applied model along with the values of its parameter; (iii) the diffusion process seeds.
So far our experiments were performed implicitly fixing all those parameters once and for all. 
Indeed, the initial set of infected nodes has a relevant impact on how a diffusive process unfold: for such reason \ndlib, with a single line of code, allows to execute several times, in parallel, the simulation of the same model over a given network topology while varying the initial infection status.

By using \texttt{ndlib.utils.multi\_runs} the initial infected nodes for each model instance can be specified either:
\begin{itemize}
	\item by the \texttt{percentage\_infected} model parameter, or
	\item explicitly through a list of $n$ sets of nodes (where n is the number of executions required)
\end{itemize}
Note that, in the first scenario \texttt{ndlib.utils.multi\_runs} nodes will be sampled independently for each model execution.
\begin{figure*}[t]
  \centering 
\includegraphics[width=0.95\columnwidth]{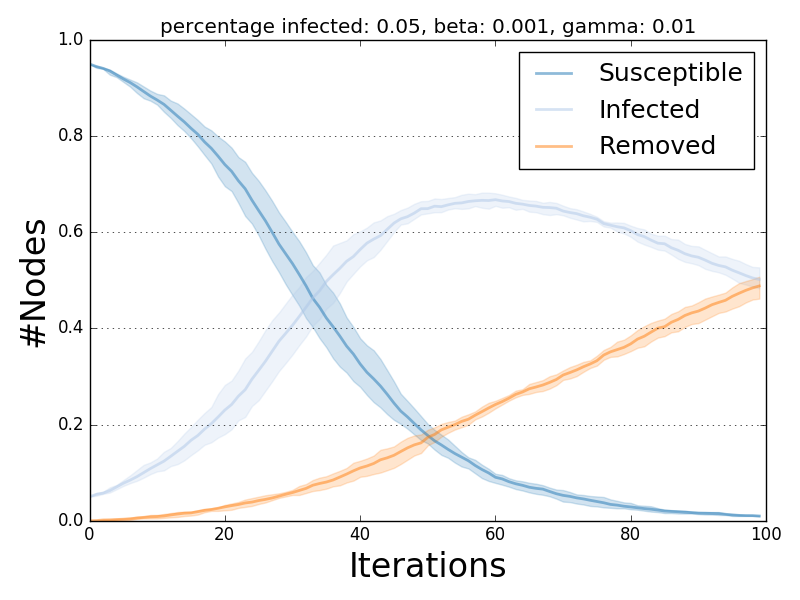}\qquad
\includegraphics[width=0.95\columnwidth]{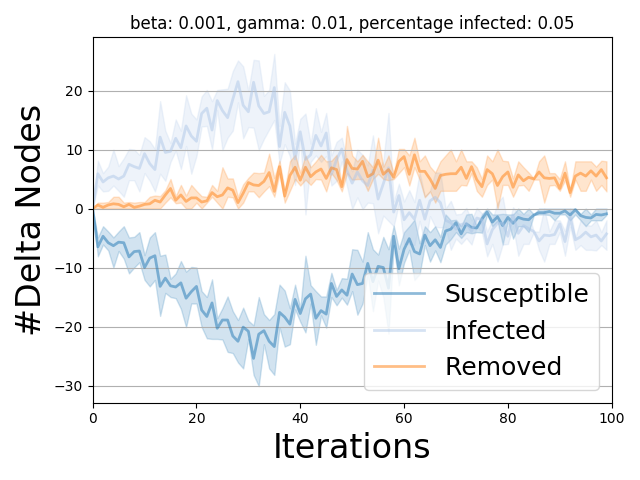}
  \caption{{\bf Model Comparison.} \ndlib\ Multiple executions for SIR model, areas identify inter-quartile ranges.}
  \label{img:multiple}
\end{figure*}

Results of parallel executions can be feed directly to all the visualization facilities exposed by \texttt{ndlib.viz}, thus producing the previously introduced plots.
In case of multiple executions, all line plots report the average trend lines along with a user-defined percentile range (as in Figure \ref{img:multiple}) -- a solution that allows for a better understanding of model stability.

\subsection{Dynamic Network Support}
\label{subsec:dynetx}
Diffusive processes can unfold on top of dynamic network topologies.
Moreover, nodes and edges dynamics deeply impact the evolution of diffusive phenomena: in some scenarios, the adoption of a temporally flattened graph topology can lead to an overestimate of the network connection resulting, for instance, in an overall speed up of the observed spreading process.
Conversely, analyzing a potentially sparse time evolving network structure can produce very different outcomes.

To allow \ndlib\ users to describe, and simulate, models on top of dynamic topologies we designed a novel dynamic network library, \dynetx,  and define a class of diffusive models \\ (grouped under \texttt{ndlib.models.dynamic}).

\dynetx, whose name stands for \emph{``Dynamic NetworkX"}, is a Python package that extends the network facilities offered by NetworkX mainly providing two novel base graph classes, namely, \texttt{DynGraph} -- dynamic undirected graphs -- and \texttt{DynDiGraph} -- dynamic directed graphs.
The library, available for Python 2.7.x and 3.x, is currently hosted on GitHub\footnote{\dynetx\ GitHub: \url{https://goo.gl/UL3JVd}}, on pypi\footnote{\dynetx\ pypi: \url{https://goo.gl/8HbS8d}} and has its online documentation on ReadTheDocs\footnote{\dynetx\ docs: \url{https://goo.gl/ngKHpY}}.

\subsubsection{\dynetx\ Rationale}
\dynetx\ provides support to temporal networks as well as to snapshot graphs through a unique interface.
The idea behind \dynetx\ is to balance two needs: (i) minimize the amount of information needed to model dynamic graphs, and (ii) maintain a \emph{pythonic} interface. 
To fulfill such goals while avoiding a complete reimplementation of a graph model we decided to extend the basilar classes offered by NetworkX for both undirected and directed graphs.
Pursuing such choice we were able also to maintain some degree of compatibility among the two libraries allowing, for instance, \dynetx\ \texttt{DynGraph} objects to be naturally converted in NetworkX \texttt{Graph} ones when flattened/sliced.

To introduce the dynamic information to, an otherwise, static model we make \emph{edges} into \emph{interactions}: to such aim we extended each edge with a list describing the temporal interval of its presence in the dynamic topology.
In our model, an \emph{interaction} among nodes $u,v \in V$ presents at timestamps $[t_1, t_2, t_3, t_5, t_6]$ is modeled as a triple $i_{u,v} = (u, v, < [t_1,t_4), [t_5,t_7))>)$	
in which the third component provides a compact view of the presence intervals, each having form $[t_i, t_j)$ with $t_i<t_j$, where the left side, $t_i$, is included in the interval while the second, $t_j$, is not. 
Such solution allows to limit the data structure size to $O(|E|*|T|)$ where $|E|$ represent the cardinality of the flattened edge set and $|T|$ the cardinality of the timestamp set.
Moreover, the provided representation does not particularly affect the standard NetworkX graph model thus limiting the overall impact on scalability.

\subsubsection{Diffusive Phenomena in Dynamic Networks.}

Following the already discussed \ndlib\ rationale, all dynamic models extend an abstract class, namely\\ \texttt{DynamicDiffusionModel} and implement three of its methods:
\begin{itemize}
	\item[i)] \texttt{\_\_init\_\_(self, graph)}, the constructor used to specify model metadata;
	\item[ii)] \texttt{execute\_snapshots()}, that describe the specific model rules to apply when the underlying topology is modeled as a Snapshot graph;
	\item[iii)] \texttt{execute\_iterations()}, that describe how to simulate the model on top of Temporal networks.
\end{itemize}  

In order to provide some examples of how classic epidemic models can be revised to handle evolving graph topologies in \ndlib\ are implemented the dynamic version (both for snapshot graphs and temporal networks) of \emph{SI}, \emph{SIS} and \emph{SIR}: namely \emph{DynSI}, \emph{DynSIS} and \emph{DynSIR}.
This latter class of models maintains the same configuration strategy and visualization facilities previously introduced.

\subsection{Extensibility: User Defined Models}
As discussed before, all the diffusion models implemented in \ndlib\ extend the same template exposed by \\ \texttt{DiffusionModel}.
The \texttt{DiffusionModel} abstract class takes care of: (i) validate the coherence and completeness of the \texttt{ModelConfig} object used to setup the simulation; (ii) initialize the simulation; (iii) define an  interface for parameter passing and model execution.

Extending the \ndlib\ library is easy. 
A novel model, identified by a Python class that extends \texttt{DiffusionModel}, must implement two methods: \\ (i) its constuctor (e.g. \texttt{\_\_init\_\_(self, graph)}), and (ii) the iteration step (e.g. \texttt{iteration()}).
The \texttt{\_\_init\_\_} method is used to provide a meta description of the model parameters (both global and node/edge specific) as well as for the node statuses.
Such model meta-data have two roles: they allows \texttt{DiffusionModel} to check the consistency of the model configuration and enable the \texttt{VisualizeDiffusion} object to customize the simulation visualizations. 
%
%
%
%
%
%
%
%
The core of each diffusion model is thus defined in its \texttt{iteration()} method.
As discussed before the entire diffusion process can be seen as an agent-based discrete-time simulation: the \texttt{iteration()} method describe the rules that decide for each \emph{agent} (i.e. a node), during a round of simulation, if it will maintain its status or change it.

The iteration step is composed of three stages: (i) collection of the actual nodes' statuses, (ii) update cycle over the nodes, and (iii) computation of the incremental updates.
Note that the first step is mandatory since we consider each iteration as atomic and we expect synchronous updates of all the nodes (e.g., the model rules must be applied to all the agents starting from the same initial configuration).
Since \ndlib\ is actively maintained, for a more detailed discussion on library extension please refer to the official documentation\footnote{Extending \ndlib: \url{https://goo.gl/ycoUQ8}}.
%
%
%
%
%
%
%
%
%
%
%

\subsection{Experiment Server}

As classical analytical tools, the simulation facilities offered by \ndlib\ are specifically designed for those users that want to run experiments on their local machine.
However, in some scenarios -- e.g., due to limited computational resources or to the rising of other particular needs -- it may be convenient to separate the machine on which the definition of the experiment is made from the one that actually executes the simulation.
To satisfy such needs, we developed a RESTful service, \ndlib-REST\footnote{\ndlib-REST: \url{https://goo.gl/c6yHcY}}, that builds upon \ndlib\ an experiment server queryable through API calls.

\subsubsection{\ndlib-REST rationale.}
The simulation web service is designed around the concept of \emph{experiment}.
An experiment, identified by a unique identifier, is composed of two entities: (i) a network and (ii) one (or more) configured models.
Experiments are used to keep track of the simulation definition, to return consecutive model iterations to the user and to store - locally on the experiment server - the current status of the diffusion process.

\begin{figure*}[t]
  \centering 
\includegraphics[width=0.9\linewidth]{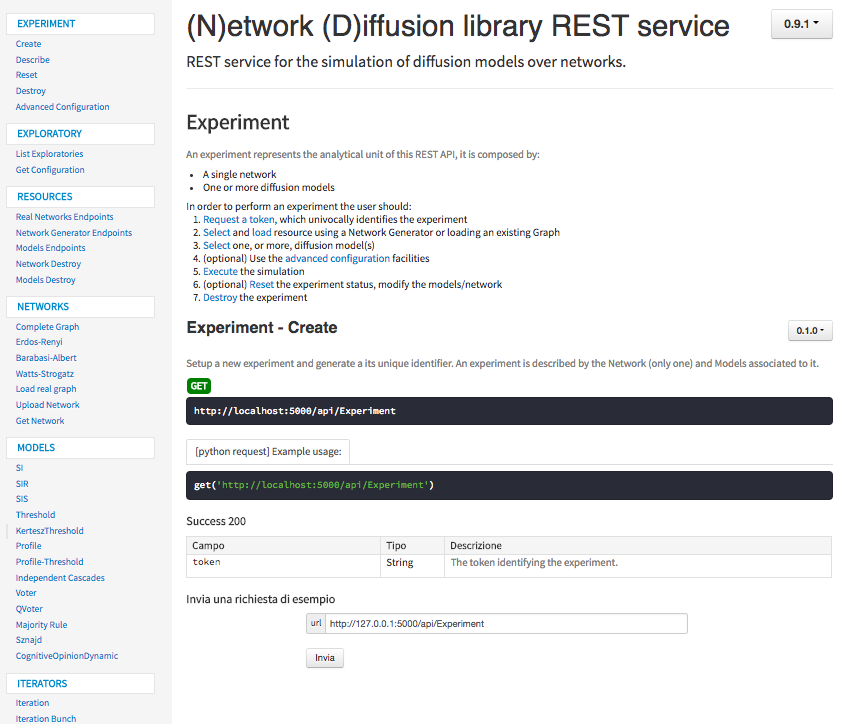} 
  \caption{{\bf Experiment Server.} \ndlib-REST documentation webpage.}\label{img:ndlib_rest}
\end{figure*}

In particular, to perform an experiment, a user must:
\begin{itemize}
	\item[1.] Request a token, which univocally identifies the experiment;
	\item[2.] Select or load a network resource;
	\item[3.] Select one, or more, diffusion model(s);
	\item[4.] (optional) Use the advanced configuration facilities to define node/edge parameters;
	\item[5.] Execute the step-by-step simulation;
	\item[6.] (optional) Reset the experiment status, modify the models/network;
	\item[7.] Destroy the experiment.
\end{itemize}
The last action, involving the destruction of the experiment, is designed to clean the serialization made by the service of the incremental experiment status.
If an experiment is not explicitly destroyed its status is removed, and the associated token invalidated, after a temporal window that can be configured by the service administrator.
\ndlib-REST is shipped as a Docker\footnote{Docker: \url{https://goo.gl/BbWz7X}} container image so to make it configuration free and easier to setup.
Moreover, the simulation server is, by default, executed within a Gunicorn\footnote{Gunicorn: \url{https://goo.gl/TFjebL}} instance allowing parallel executions of multiple experiments at the same time.
\ndlib-REST is built using Flask\footnote{Flask: \url{https://goo.gl/qQhcBn}} and offers a standard online documentation page (shown in Figure \ref{img:ndlib_rest}) that can also be directly used to test the exposed endpoints both configuring and running experiments.

\subsubsection{Querying \ndlib-REST}
As a standard for REST services, all the calls made to \ndlib-REST  endpoints generate JSON responses.
The API (Application Programming Interface) of the simulation service is organized into six categories so to provide a logic separation among all the exposed resources. 
In particular, such categories are:
\begin{itemize}
	\item {\em Experiments}: endpoints in this category allow to create, destroy, configure, reset and describe experiments;
	\item {\em Exploratories}: endpoints in this category allow to load predefined scenarios (e.g., specific networks/models with explicit initial configuration);
	\item {\em Resources}: endpoints in this category allow to query the system to dynamically discover the endpoints (and their descriptions) defined within the system;
	\item {\em Networks}: endpoints in this category handle a load of network data as well as the generation of synthetic graphs (Barabasi-Albert, Erd\"os-Renyi, Watts-Strogatz, \dots); 
	\item {\em Models}: endpoints in this category expose the \ndlib\ models;
	\item {\em Iterators}: endpoints in this category expose the step-by-step and iteration bunch facilities needed to run the simulation. 
\end{itemize}
Since the simulation service allows to attach multiple diffusion models to the same experiment both the single iteration and the iteration bunch endpoints expose an additional parameter that allows the user to select the models for which the call was invoked. 
By default, when such parameter is not specified, all the models are executed and their incremental statuses returned.
A particular class of endpoints is the {\emph Exploratories} one.
Such endpoints are used to define the access to pre-set diffusion scenarios.
Using such facilities the owner of the simulation server can describe, beforehand, specific scenarios (network, initial node states\dots), package them and make them available to the service users.
From an educational point of view such mechanism can be used, for instance, by professors to design emblematic diffusion scenarios (composed by both network and initial nodes/edges statuses) so to let the students explore their impact on specific models configurations (e.g., to analyze the role of weak-ties and/or community structures).
To provide a simplified interface to query the \ndlib-REST service, we defined a Python wrapper, shipped along with the experimental server.

\begin{figure*}[ht]
  \centering 
\includegraphics[width=1\textwidth]{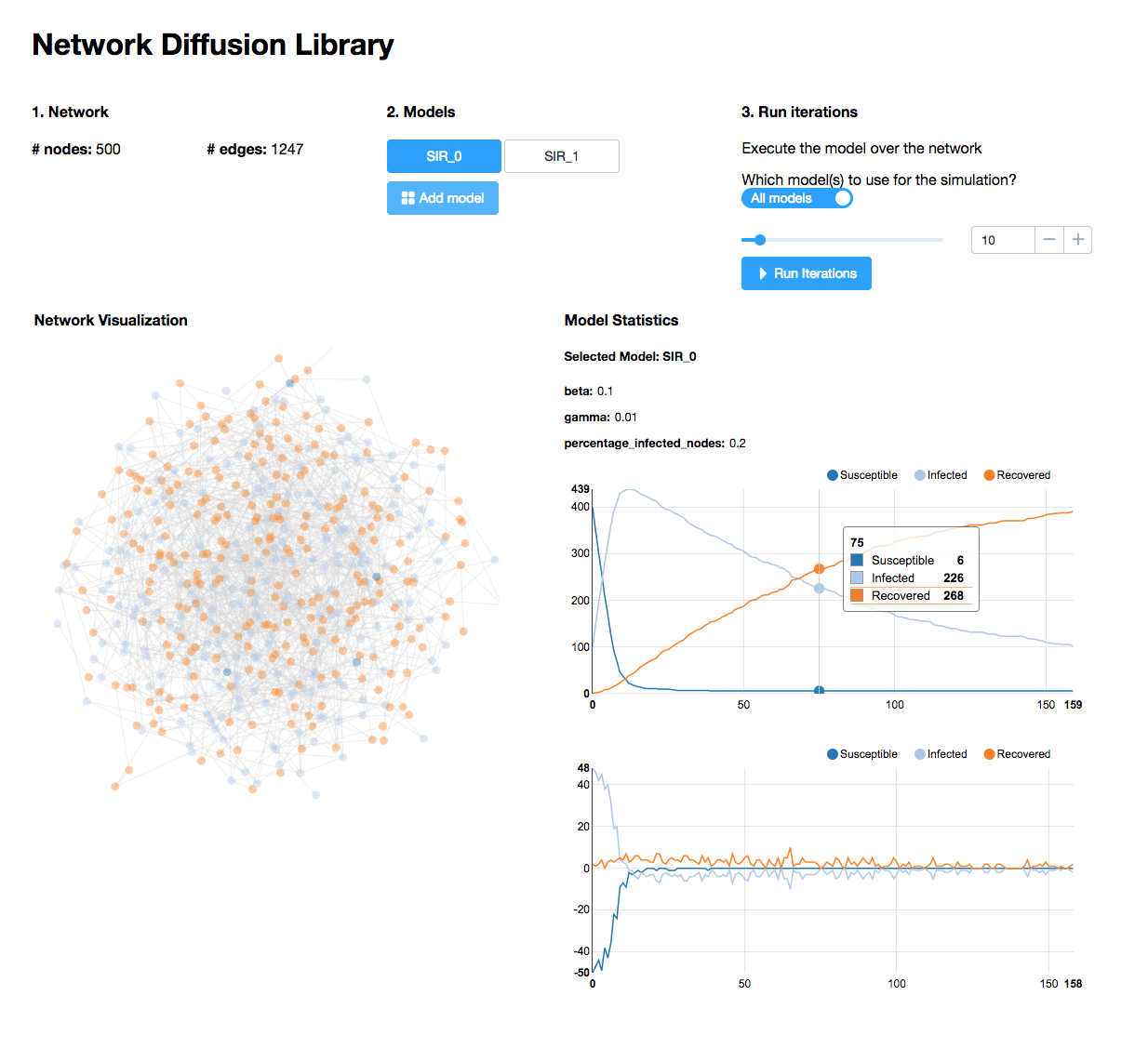}  \caption{{\bf Web Platform.} \ndlib\ Visualization Framework appearance during a simulation. The top toolbar presents a schematic workflow to execute a simulation. The left view presents a visualization of the status of each node. The right part presents a synthetic visualization of properties of the simulation. Mouse interaction allows the user to select a specific time instant of the simulation to update all the other views accordingly}\label{img:viz-framework}
\end{figure*}

\subsection{Web Platform}
Finally, upon the \ndlib-REST service, is designed a visualization platform\footnote{Available, currently as a separate branch, at: \url{https://goo.gl/tYi48o}}.
\ndlib-Viz aims to make non-technicians able to design, configure and run epidemic simulations, thus removing the barriers introduced by the usual requirements of programming language knowledge.
Indeed, apart from the usual research-oriented audience, we developed \ndlib-Viz to support students and facilitate teachers to introduce epidemic models: in order to better support the educational nature of the proposed platform and collect user feedbacks, we currently employ \ndlib-Viz in a Social Analysis course of the Computer Science Master Degree at the University of Pisa, Italy.
The platform itself is a web application: it can be executed on a local as well as on a remote \ndlib-REST installation. 
The visual interface guides the user to follow the \ndlib-REST workflow through a toolbar on top of the page with a schematic representation of the expected steps (see Figure~\ref{img:viz-framework}). 
As a first step, the user should create an experiment and a network. 
The parameters to import or generate the network are exposed via a web form, also providing suggestions and error checking for the parameters entered by the user. 
Once the network has been created, it is rendered on the screen in a viewport. 
At the second step, the user may create one or more diffusion models to attach to the network. 
Each model is simulated according to the specifics of \ndlib-REST. 
The simulation is handled by the user in the third step, where she can choose the number of iteration to execute. 
The time discrete-time simulation is presented to the user by mean of a set of linked displays \cite{Newton197859}, where the selection and interaction performed on a view are propagated to the other views. 
The main view is the visualization of the network, showing all the nodes and their links. 
Each node is assigned a color to represent its status in a specific time instant of the simulation. 
The choice of the time instant to visualize is determined by the user, by means of selection of the mouse on the linked displays.
The bottom part of the interface shows the result of simulation for each model. 
An aggregated visualization of each model is presented in a block containing the reference to the model and its parameters, and two charts to show \texttt{DiffusionTrend} and \texttt{DiffusionPrevalence} plots, like those presented in Section~\ref{sec:ndlibenv}. 
Exploiting the web interface, the plots are interactive. 
Exploring the plots with the mouse the user may receive additional information on specific time instant (see tooltip box in the example in Figure \ref{img:viz-framework}). The selection on the plot is directly linked with the block of network visualization, showing the node statuses in the corresponding time instant.

\subsubsection{Architecture}
The Visualization Framework is a single page web application implemented using Javascript and HTML 5. 
The decoupling of the simulation engine and the visual interface allow us to exploit  modern browsers to provide an efficient environment for visualization of models and interactions. 
The structure and layout of the page are managed with Bootstrap\footnote{Bootstrap: \url{https://goo.gl/8xhPdm}}. 
The business logic and visualization of graphical widgets are implemented in D3.js\footnote{D3.js: \url{https://goo.gl/mPrn3i}}. 
Nodes and edges of the networks are drawn using the Force Layout library provided by the D3 library. 
The network visualization is implemented using Canvas object provided by standard HTML5. 
This allows a very efficient update of the network view. 
The charts showing the Diffusion Trend and Prevalence (presented in Section \ref{sec:ndlibenv}) are created using NVD3 library\footnote{NVD3: \url{https://goo.gl/YW87eA}}.

The Visualization Framework is implemented using a Model-Control-View (MCV) design pattern. 
The model is managed by a central component that implements a REST API client that handle the status of the experiment. 
When the user interacts with one of the views (charts, network layout, toolbar) the controller notifies the model to update the experiment. 
Each interaction with the visual interface is managed by the model component that centralizes all the communications with the REST server. 
The calls to the server are executed asynchronously and the component updates the visual interface as soon as a response arrives from the server. 

\section{Comparative Analysis}
\label{sec:tools}
When it comes to model and study complex networks, as well as the dynamic  phenomena they are used to model, several resources are available to students, programmers and researchers.
In this section, we propose a succinct review of the most recent/used libraries and simulators, \ref{sub:sim}.
Indeed, our analysis does not aim to cover all the existing resources but only the ones that, in our opinion, provide interesting facilities. 
Moreover, in \ref{sub:comparison} we compare \ndlib\ with some of the identified libraries both qualitatively -- by discussing their features --, and quantitatively -- by compare scalability and running time for a same diffusion model.

\subsection{\ndlib\ Competitors}
\label{sub:sim}

Indeed, \ndlib\ is not the first library designed to model, simulate and study diffusive dynamics on complex networks.
To better frame our library within the set of existing analytical tools we identified the following competitors. 
\begin{itemize}
	\item {\bf Epigrass}\footnote{Epigrass: \url{https://goo.gl/yjbLRk}} is a platform for epidemiological simulation and analysis on geographical networks \cite{coelho2008epigrass}. Epigrass is entirely written in the Python language and uses the NetworkX library to manage the networks. It provides epidemic models, such as SIR, SIS, SEIR and SEIS and some variations of these models. 
	\item {\bf GEMF-sim}\footnote{GEMF-sim: \url{https://goo.gl/kcGs6R}} is a software tool that implements the generalized formulation of the epidemic spreading problem and the related modeling solution \cite{sahneh2017gemfsim}. It is available in popular scientific programming platforms such as C, Python, R, and MATLAB. The models implemented covers the most common epidemic ones. It can be applied to spreading processes with various node states and contact layers; it allows users to incorporate mitigation strategies, such as the adoption of preventive behaviors and contact tracing in the study of disease spreading.
	\item {\bf Nepidemix}\footnote{Nepidemix: \url{https://goo.gl/M8rEGM}} is a suite  tailored to programmatically describe simulation of complex processes on networks \cite{ahrenberg2016nepidemix}. Nepidemix was developed by members of the IMPACT-HIV group; it is written in Python 2 and uses the module NetworkX to manage the network structure. It currently provides three epidemic models: SIS, SIR and SIJR. It automates common diffusion simulation steps allowing the programmer to build a network according to some specifics and to run on top of it a set of epidemic processes for a specified number of iterations.  Moreover, Nepidemix allows during simulation to save incremental results such as disease prevalence and state transitions.
	\item {\bf EoN}\footnote{EoN: \url{https://goo.gl/cuArFP}} is another Python library dedicated to the simulation of diffusive models. EoN is designed to study the spread of SIS and SIR diseases in networks \cite{211986}. It is composed of two sets of algorithms: the first set that deals with simulation of epidemics on networks (SIR and SIS) and the second designed to provide solutions of systems of equations. Also, this package is built on top of NetworkX graph structures.
	\item {\bf Epydemic}\footnote{Epydemic: \url{https://goo.gl/PrPHh4}} is a library for performing simulations for two epidemic spreading processes (SIS and SIR), simulated over networks represented using NetworkX. It provides the basic simulation machinery for performing epidemic simulations under two different simulation regimes: synchronous simulation in which time proceeds in discrete time intervals, and stochastic simulations.
	\item {\bf ComplexNetworkSim}\footnote{ComplexNetworkSim: \url{https://goo.gl/nczJTH}} is a Python package for the simulation of agents connected in a complex network. The framework is designed for users having computer science background; it can create a virtual complex network with agents that interact with each other. This project is not limited to a static network but also takes into account temporal networks, where processes can dynamically change the underlying network structure over time. Currently, it provides two types of epidemic models: SIS and SIR.
	\item {\bf Nxsim}\footnote{Nxsim: \url{https://goo.gl/U2rDvv}} is a Python package for simulating agents connected by any type of network using SimPy and NetworkX in Python 3.4. This project is a fork of the previous package, ComplexNetworkSim. 
	\item {\bf EpiModel}\footnote{EpiModel: \url{https://goo.gl/g9RRCM}} is one of the most famous package written in R \cite{epimodel}. EpiModel provides facilities for build, solve, and plot mathematical models of infectious disease. It currently provides functionality for three classes of epidemic models -- Deterministic Compartmental Models, Stochastic Individual Contact Models and Stochastic Network Models -- and  three types of infectious disease can be simulated upon them: SI, SIR, SIS. This package is built on top of Igraph network structures. EpiModel allows generating visual summaries for the execution of epidemic models; it provides plotting facilities to show the means and standard deviations across multiple simulations while varying the initial infection status. It also includes a web-based visual application for simulating epidemics\footnote{EpiModel-viz: \url{https://goo.gl/13Z5mj}}.
	\item {\bf RECON}\footnote{RECON: \url{https://goo.gl/eYMDqh}}, the R Epidemic Consortium, assembles a group of international experts in infectious disease modeling, Public Health, and software development to create the next generation of tools for disease outbreak analysis using the R software. The project includes an R package to compute, visualize and model disease incidents.
	\item {\bf Sisspread}\footnote{Sisspread: \url{https://goo.gl/LSWsUh}} allows simulating the dynamics of an hypothetic infectious disease within a contact network of connected people. It was written in C and it implements three classical schemes of disease evolution (SI, SIS and SIR), and may evaluate the spreading on different contact networks topologies (random homogeneous, scale-free, small-world), and also on user-provided networks. 
	\item {\bf GLEaMviz}\footnote{GLEaMviz: \url{https://goo.gl/kTftxZ}} is a publicly available software that simulates the spread of emerging human-to-human infectious diseases on world scale \cite{van2011gleamviz}. The GLEaMviz framework is composed of three components: the client application, the proxy middleware, and the simulation engine. The simulations it defines combine real-world data on populations and human mobility with elaborate stochastic models of disease transmission to simulate disease spread on the global scale. As output, it provides a dynamic map and several charts describing the geo-temporal evolution of the disease.
\end{itemize}

\noindent The previously listed resources are designed to allow the final user to simulate epidemic models in networked contexts following different rationales.
However, due to the interdisciplinary nature of the specific problem tackled, there are also a lot of single model libraries aimed to simulate a specific disease or, conversely general simulation tools exposing few ad-hoc epidemic models.
Examples of such categories are:
\begin{itemize}
	\item {\bf NetLogo}\footnote{NetLogo: \url{https://goo.gl/82zAoc}} a programmable modelling environment for simulating natural and social phenomena. It was developed by Uri Wilensky in 1999 \cite{wilensky1999netlogo} and has been in continuous development ever since at the  ``Center for Connected Learning and Computer-Based Modeling". It is particularly well suited for modeling complex systems that evolve over time describing them as agent-based processes. NetLogo enables users to run a predefined set of simulations and play with their parameters, exploring their behaviors under various conditions. 
	\item {\bf System Sciences}\footnote{System Sciences: \url{https://goo.gl/wH5kJn}} is an online project created by the ``Institute of Systems Sciences, Innovation and Sustainability Research" at the University of Graz whose aim is to design an interactive electronic textbook for systems sciences based on software applications for tablet computers. In the disease spreading section offered by this tool, the user can  choose a network from a set of classical network models (random, small world, scale free and complete network) and then fix the parameter of the SIR model (the only one implemented so far). 
	\item {\bf FRED}\footnote{FRED: \url{https://goo.gl/JwUx7k}} (A Framework for Reconstructing Epidemiological Dynamics) is an open source modeling system developed by the ``Pitt Public Health Dynamics Laboratory" in collaboration with the ``Pittsburgh Supercomputing Center and the School of Computer Science" at Carnegie Mellon University \cite{grefenstette2013fred}. FRED supports research on the dynamics of infectious disease epidemics and the interacting effects of mitigation strategies, viral evolution and personal health behavior. The system uses agent-based modeling based on census synthetic populations data that capture the demographic and geographic distributions of the population. FRED epidemic models are currently available for every state and country in the United States, and for selected international locations.
	\item {\bf FluTE}\footnote{FluTE: \url{https://goo.gl/VEBshU}} is an individual-based model capable of simulating the spread of influenza across major \\metropolitan areas or the continental United States \cite{chao2010flute}. It can simulate several intervention strategies, and these can either change the transmission characteristics of influenza (e.g., vaccination) or change the contact probabilities between individuals (e.g., social distancing). It is written in C/C++.
	\item {\bf Malaria Tool}\footnote{Malaria: \url{https://goo.gl/yyStf9}} is a user-interface to a combined intervention model for malaria which was developed by Imperial College London as part of the Vaccine Modeling Initiative. 
	\item {\bf EpiFire}\footnote{EpiFire: \url{https://goo.gl/W5QLb1}} is a fast C++ applications programming interface for simulating the spread of epidemics on contact networks.
	\item {\bf Measles Virus}\footnote{Measles: \url{https://goo.gl/R1aM1N}} is an application written in Python and Matlab for the simulation of the spread of the measles virus \cite{word2010estimating}.
\end{itemize}

\begin{table*}[!t]\centering
{
\scriptsize 
\renewcommand{\arraystretch}{1.5}
    \begin{tabular}{|l|m{0.9cm}|m{0.7cm}|m{0.7cm}|m{0.7cm}|m{0.7cm}|m{0.9cm}|m{1.3cm}|l|m{1.1cm}|m{0.9cm}|m{1.1cm}|}        
        \hline
        {\bf Name}  & {\bf Lang.} & {\bf Epi.} & {\bf Op. Dyn.} & {\bf Viz.} & {\bf Dyn. Net.} & {\bf Exp. Server} & {\bf Viz. Platform} & {\bf Ext.} & {\bf Net. Model} & {\bf Active} & {\bf License}\\
        \hline
        \ndlib & Python & \checkmark (14) & \checkmark (5) & \checkmark & \checkmark & \checkmark & \checkmark & \checkmark & NetworkX DyNetX & \checkmark & BSD \\
        epigrass & Python & \checkmark (8) & & \checkmark & & & \checkmark & \checkmark & NetworkX & \checkmark & GPL \\
        GEMFsim & Python & \checkmark (4) & &  & & & & \checkmark & NetworkX & \checkmark & - \\
        Nepidemix & Python & \checkmark (3) & & & & & \checkmark & \checkmark & NetworkX & \checkmark & BSD\\
        EoN & Python & \checkmark (3)  & & & & & & & NetworkX & \checkmark & MIT \\
        epydemic & Python & \checkmark & & & & & & & NetworkX & \checkmark & GPL \\
        ComplexNetworkSim & Python & \checkmark (3) & & \checkmark & \checkmark & & & \checkmark & NetworkX & & BSD \\
        nxsim  & Python & \checkmark (3) & & \checkmark & \checkmark & & & \checkmark & NetworkX & & Apache\\
        
        EpiModel & R & \checkmark (3) & & \checkmark & & & \checkmark & \checkmark & igraph & \checkmark & GPL \\
        RECON & R & \checkmark & & \checkmark & & & \checkmark & \checkmark & adhoc & \checkmark & Various\\
        sisspread & C & \checkmark (3) & &  & & & & & adhoc & \checkmark & GPL\\
        
        \hline
       	 GLEaMviz & C++ Python & \checkmark & & \checkmark & & \checkmark & \checkmark & \checkmark & adhoc & \checkmark & SaaS \\
     \hline
    \end{tabular}
    \renewcommand{\arraystretch}{1}
    }
    \caption{{\bf Diffusion Libraries and Tools.} A qualitative comparision of network diffusion related libraries and tools. Within brackets the number of diffusion models, if any, natively implemented within each library.}
    \label{tab:comparision}
\end{table*}

\subsection{Comparisions}
\label{sub:comparison}
Comparing different libraries is not an easy task.
Indeed, the choice of underlying technologies, programming languages, audiences as well as final aims profoundly shape a suite of analytical tools.
In the following we selected a subset of the previously introduced frameworks and proposed a two level comparison covering both \emph{qualitative}, \ref{subsub:qual}, and \emph{quantitative}, \ref{subsub:quant}, aspects.  

\subsubsection{Qualitative}
\label{subsub:qual}
In order to qualitatively compare the selected libraries we identified the following set of features:
\begin{itemize}
	\item \emph{Programming Language.} The choice of a programming language deeply affects library performances. Moreover, in some cases, it implicitly defines to some extent the prototypical user of the library itself. 
	\item \emph{Epidemic Models.} Whether the library is shipped with native supports for simulating epidemic models. 
	\item \emph{Opinion Dynamic Models.} Whether the library is shipped with native supports for simulating opinion dynamic models. 
	\item \emph{Embedded Visualization Facilities.} Whether the library is shipped with native visualization facilities. 
	\item \emph{Dynamic Network Support.} Whether the library is shipped with native support for dynamic networks. 
	\item \emph{Experiment Server.} Whether the library allows running experiment server out of the box. 
	\item \emph{Visual Platform.} Whether the library is shipped with a Visual Platform for non programmers. 
	\item \emph{Extensibility.} Whether the library allows the definition of novel diffusion models. 
	\item \emph{Network Model.} The underlying library used to model graph topologies (this library affects scalability). 
	\item \emph{Project Status.} Whether the project is currently developed or if its support ceased.
	\item \emph{Library License.} Whether the library is open source or not.
\end{itemize}  
Table \ref{tab:comparision} reports a characterization of the selected libraries from the perspective of the identified features.
We can observe that most of the identified libraries are written in Python and exploit the NetworkX package as underlying network modeling framework.
Such trend confirms the fact that Python (as well as R) represents nowadays one of the main languages in the data science community.
Moreover, most of the analyzed libraries are actively maintained at the moment of our survey and, with few exceptions, are released with licenses associated with open software development.
Moving to library related characteristics, we observe a clear pattern: all libraries offer out of the box support for epidemic models (often providing a subset of classic compartmental models) while none, apart for \ndlib\ natively support opinion dynamic ones.
Moreover, only three libraries support dynamic network topologies to some extent and among them \ndlib\ is the only active project.
From our analysis emerges that half of the libraries taken into account provides predefined visualization facilities or visual platforms tailored to support the analysis of non developers. 
Another interesting result is that almost all the tools identified provide some sort of extensibility, allowing their final users to implement novel diffusive models.
Finally, we can observe how only \ndlib\ and GLEaMviz offer an experiment server to their users.
However, GLEaMviz is an outsider among the compared tools: it only comes with a visual interface and does not allow its users to  use it as a programming library. 
Moreover, to the best of our knowledge, for such tool, only a client platform is shipped to the final users since all the experiments are run on a proprietary experiment server not publicly released.

\subsubsection{Quantitative}
\label{subsub:quant}
Two strategies can be pursued to produce a quantitative comparison of different libraries: analyzing their scalability, comparing running times of a common model they implement.
Indeed, such kind of evaluations cannot be considered exhaustive due to the different nature/aim/set of models implemented by the compared libraries and are deeply tied to both library design an programming language used.
Moreover, since \ndlib, as well as most of the identified libraries, relies on the external implementation of graph structure evaluating its scalability does not allow to make any valuable conclusion on the functionality it implements.
Indeed, graph size upper bounds will be the ones of the network model adopted thus producing a benchmark not of diffusion libraries but of network ones.

Regarding the execution running times, we decided to compare only a subset of the libraries listed in Table \ref{tab:comparision}.
In particular, we focused our attention on: \ndlib\, ComplexNetworkSim, Nepidemix and EpiModel.
The main reasons behind such reduced selection are the following:
\begin{itemize}
	\item \emph{Available Documentation.} We privileged all those libraries having a well organized user oriented documentation. In our opinion, an analytical project should provide support to its final user: otherwise, in case of out of date/inconsistent/missing documentation, the library it is not a suitable choice for those who need to focus on experiment definition and execution.
	\item \emph{Library Organization.} Among the libraries filtered using the previous criterion, we selected those libraries that are not merely a collection of models but, conversely, expose a common schema for approach definition and instantiation.
\end{itemize}
To provide a fair comparison we executed for each library the same model, SIR, with the same parameter settings, on the same class of network while varying the number of nodes.
All the tests were run on a workstation with an Intel Core i7-5820K 12 core processors at 3.30Ghz, with 32GB of RAM, running Linux 4.4.0-93. 
The execution times, reported in Table \ref{tab:runtime}, refer to the execution of 25 iterations of the selected model without taking into account the graph creation step.
Our results underline that, in the selected scenario, \ndlib\ is capable of outperforming its direct competitors -- namely, ComplexNetworkSim and Nepidemix. 
Indeed, it always completes its execution one order of magnitude faster than the other Phyton, NetworkX based, diffusion libraries.
EpiModel, conversely from the other competitors, is implemented in R and leverages the iGraph library (written in C) as graph modeling framework.
Although not been a Python library we included EpiModel in this quantitative analysis since, so far, it represents one of the most widespread libraries for diffusion analysis.
The running time comparison favors EpiModel: however, such result -- likely due to the different data structure adopted -- does not underline a wide gap between the two libraries.

\begin{table}[!t]\centering
{
\scriptsize 
\renewcommand{\arraystretch}{1.5}
    \begin{tabular}{|l|m{0.9cm}|m{0.9cm}|m{0.9cm}|m{0.9cm}|}        
        \hline
       \multirow{2}{*}{\bf Library} & \multicolumn{4}{c|}{\bf Graph Size (nodes)}\\
        \cline{2-5}
         & {\bf 10$^3$} & {\bf 10$^4$} & {\bf 10$^5$} & {\bf 10$^6$} \\
        \hline
        \ndlib & 0.060s & 0.655s & 7.554s & 90.443s \\
        ComplexNetworkSim & 0.264s & 3.152s & 43.145s & 576.072s  \\
        Nepidemix & 0.283s & 3.241s & 43.190s & 525.768s\\
        \hline
        EpiModel & 0.025s & 0.141s & 2.289s & 45.725s\\

     \hline
    \end{tabular}
    \renewcommand{\arraystretch}{1}
    }
    \caption{{\bf Runtimes Comparison}. SIR model (parameters: $\beta=0.001$, $\gamma=0.01$), initial infected 5\%, number of iterations 25, network model Barabasi-Albert graph. The compared libraries are organized by programming language.}
    \label{tab:runtime}
\end{table}

\section{Conclusions and Future Works}
\label{sec:conclusion}
In this paper, we introduced \ndlib\, a modular framework designed to provide easy access to network diffusion simulation models to a broad user base.
\ndlib\ aims to support the analytical needs of a heterogeneous audience: in particular, the proposed framework is composed of a standalone library, a RESTful simulation service, and a visualization tool.
\dynetx\ as well as \ndlib\ are released as free-software under a BSD-2-Clause license.
The number of implemented models (see Appendix) as well as its extensibility make \ndlib\ a valid solution for those users that need to compare different diffusive schema other than the classical compartmental ones (SI, SIS, SIR). 

Moreover, \ndlib\ is among the few libraries that allows the definition and analysis of diffusive phenomena unfolding over dynamic network topologies: a peculiarity make possible thanks to the integration with \dynetx\ a novel library we designed to model time evolving graphs. 

\dynetx\ is among the first pure python library providing dynamic network modeling facilities. 
Indeed, its ease of use and integration with NetworkX make \dynetx\ a valuable project that can be embedded with limited complexity within existing SNA tools, as \ndlib. 

Nowadays, several tools are available to students, developers and researchers to simulate and study dynamics of and on complex networks, however \ndlib\ is among the first that offer multiple interfaces specifically designed to solve specific use case.
Moreover, among the reviewed and compared tools, only the EpiModel R package seems able to offer a comparable diffusion analytics framework in terms of library, execution time and visualization facilities. 
However, \ndlib\ is tailored for a wider audience making easy to set up remote experimental services.
Such characteristic makes our library applicable to several analytical contexts: from educational purposes (e.g., employing \ndlib-Viz), diffusion model definition and testing (e.g., leveraging its core library), to the definition of a remote experiment on massive networks (e.g., setting up an experimental server on a dedicated machine).

The proposed library aims to provide a reasonable tradeoff between ease of use and efficiency.
Since \ndlib\ leverage NetworkX data structures its scalability and efficiency is strictly tied to the ones expressed by such widespread python package.
Indeed, several graph libraries are able to handle bigger networks than NetworkX (for instance, igraph, snap\footnote{SNAP library: \url{https://goo.gl/ZYrnH9}} \cite{leskovec2016snap}, boost\footnote{Boost: \url{https://goo.gl/9xKAAR}} as well as graph-tool\footnote{graph-tool: \url{https://goo.gl/uUW5kq}} -- all of them leveraging C/C++ optimized data structure), however, since our aim is to reach a widespread, diversified, audience we identify in such library the perfect founding stone for our framework.
 
As future work, we plan to further increase the number of models implemented in the \ndlib\ and to extend the analytical methods of \dynetx.
Moreover, \ndlib-REST is in the process of being integrated into a visual simulation platform developed within the CIMPLEX H2020 EU project.

\begin{acknowledgements}
This work is funded by the European Community's H2020 Program under the funding scheme \\ ``FETPROACT-1-2014: Global Systems Science (GSS)'', grant agreement \# 641191 CIMPLEX ``Bringing CItizens, Models and Data together in Participatory, Interactive SociaL EXploratories''\footnote{CIMPLEX: \url{https://www.cimplex-project.eu}}.
This work is supported by the European Community's H2020 Program under the scheme ``INFRAIA-1-2014-2015: Research Infrastructures", grant agreement \#654024 \emph{``SoBigData: Social Mining \& Big Data Ecosystem"}\footnote{SoBigData: \url{http://www.sobigdata.eu}}.
\\ \ \\
{\bf Conflict of Interest.} On behalf of all authors, the corresponding author states that there is no conflict of interest
\end{acknowledgements}

\clearpage
\appendix
\section*{Appendix: Diffusion Methods implemented in \ndlib}

\label{sec:appendix}

\ndlib\ exposes several network diffusion models, covering both  epidemic approaches as well as and opinion dynamics.
In particular, the actual release of the library (v3.0.0) implements the following algorithms:

\section{Static Epidemic Models.}

{\bf SI}: this model was introduced in 1927 by Kermack \cite{wkermack27}.
In the SI model, during the course of an epidemics, a node is allowed to change its status only from Susceptible (S) to Infected (I).
SI assumes that if, during a generic iteration, a susceptible node comes into contact with an infected one, it becomes infected with probability $\beta$: once a node becomes infected, it stays infected (the only transition is $S\rightarrow I$).\\		

\noindent{\bf SIR}: this model was still introduced in 1927 by Kermack \cite{wkermack27}.
In the SIR model, during the course of an epidemics, a node is allowed to change its status from Susceptible (S) to Infected (I), then to Removed (R).
SIR assumes that if, during a generic iteration, a susceptible node comes into contact with an infected one, it becomes infected with probability $\beta$, then it can be switch to removed with probability $\gamma$ (the only transition allowed are $S\rightarrow I\rightarrow R$).\\	      

\noindent{\bf SIS}: as SIR, the SIS model is a variation of the SI model introduced in \cite{wkermack27}.
The model assumes that if, during a generic iteration, a susceptible node comes into contact with an infected one, it becomes infected with probability $\beta$, then it can be switch again to susceptible with probability $\lambda$ (the only transition allowed are $S\rightarrow I\rightarrow S$).\\	

\noindent{\bf SEIS}: as the previous models, the SEIS is a variation of the SI model.
For many infections, there is a significant incubation period during which the individual has been infected but is not yet infectious themselves. 
During this period the individual is in status exposed (E).
SEIS assumes that if, during a generic iteration, a susceptible node comes into contact with an infected one, it switch to exposed with probability $\beta$, then it become infected with probability $\epsilon$ and then it can be switch again to susceptible with probability $\lambda$ (the only transition allowed are $S\rightarrow E \rightarrow I\rightarrow S$).\\	      

\noindent{\bf SEIR}: as the SEIS models, the SEIR takes into consideration the incubation period, considering the status exposed (E).
SEIR assumes that if, during a generic iteration, a susceptible node comes into contact with an infected one, it switch to exposed with probability $\beta$, then it become infected with probability $\epsilon$ and then it can be switch to removed with probability $\gamma$ (the only transition allowed are $S\rightarrow E \rightarrow I\rightarrow R$).\\	      

\noindent{\bf SWIR}: this model has four states: Susceptible (S), Infected (I), Recovered (R) and Weakened (W). Besides the usual transaction $S\rightarrow I\rightarrow R$, we have also the transaction $S\rightarrow W \rightarrow I\rightarrow R$.
At time stamp n, a node in state $I$ is selected and the state of all its neighbors are checked one by one. 
If the state of a neighbor is $S$, then this state changes either i) to $I$ whit probability $\kappa$ or ii) to $W$ with likelihood $\mu$.
If the state of a neighbor is $W$, with probability $\nu$ its state changes in $I$.
Then we repeat for all node in the state $I$ the process and the state for all these nodes become $R$.\\

\noindent{\bf Threshold}: this model was introduced in 1978 by Granovetter \cite{granovetter1978threshold}.
In the Threshold model during an epidemics, a node has two distinct and mutually exclusive behavioral alternatives, e.g., it can adopt or not a given behavior, participate or not participate in a riot.
Node’s individual decision depends on the percentage of its neighbors have made the same choice, thus imposing a threshold.
The model works as follows: each node starts with its own threshold $\tau$ and status (infected or susceptible). 
During the iteration $t$ every node is observed: iff the percentage of its neighbors that were infected at time $t-1$ is grater than its threshold it becomes infected as well.\\	      

\noindent{\bf Kertesz Threshold}: this model was introduced in 2015 by Ruan et al. \cite{kerteszThreshold} and it is an extension of the Watts threshold model \cite{Watts30042002}.
The authors extend the classical model introducing a density $r$ of blocked nodes -– nodes which are immune to social influence -– and a probability of spontaneous adoption $p$ to capture external influence.
Thus, the model distinguishes three kinds of node: Blocked (B), Susceptible (S) and Adopting (A). 
A node can adopt either under its neighbors’ influence or due to endogenous effects.\\	      

\noindent{\bf Independent Cascades}: this model was introduced by Kempe et all in 2003 \cite{Kempe}.
The Independent Cascades model starts with an initial set of active nodes $A_0$: the diffusive process unfolds in discrete steps according to the following randomized rule:
\begin{itemize}
	\item When node $v$ becomes active in step $t$, it is given a single chance to activate each currently inactive neighbor $w$; it succeeds with a probability $p_{v,w}$.
	\item If $w$ has multiple newly activated neighbors, their attempts are sequenced in an arbitrary order.
	\item If $v$ succeeds, then w will become active in step $t+1$; but whether or not $v$ succeeds, it cannot make any further attempts to activate $w$ in subsequent rounds.
\end{itemize}
The process runs until no more activations are possible.\\

\noindent{\bf Node Profile}: this model is a variation of the Threshold one, introduced in \cite{MRPG17}. It assumes that the diffusion process is only apparent; each node decides to adopt or not a given behavior -– once known its existence –- only on the basis of its own interests.
In this scenario the peer pressure is completely ruled out from the overall model: it is not important how many of its neighbors have adopted a specific behavior, if the node does not like it, it will not change its interests.
Each node has its own profile describing how many it is likely to accept a behavior similar to the one that is currently spreading.
The diffusion process starts from a set of nodes that have already adopted a given behavior $H$: for each of the susceptible nodes’ in the neighborhood of a node $u$ that has already adopted $H$, an unbalanced coin is flipped, the unbalance given by the personal profile of the susceptible node; if a positive result is obtained the susceptible node will adopt the behaviour.\\  

\noindent{\bf Node Profile-Threshold}: this model, still extension of the Threshold one \cite{MRPG17}, assumes the existence of node profiles that act as preferential schemas for individual tastes but relax the constraints imposed by the Profile model by letting nodes influenceable via peer pressure mechanisms.
The peer pressure is modeled with a threshold.
The diffusion process starts from a set of nodes that have already adopted a given behavior $H$: for each of the susceptible nodes an unbalanced coin is flipped if the percentage of its neighbors that are already infected exceeds its threshold. 
As in the Profile Model the coin unbalance is given by the personal profile of the susceptible node; if a positive result is obtained the susceptible node will adopt the behavior.

\section{Dynamic Epidemic Models.}
\ndlib\ (starting from v3.0.0) implements dynamic network version of classic compartmental models (SI, SIS, SIR) \cite{MRPG17a} leveraging \dynetx\ graph structures.
Such models, here shortly described, are defined for snapshot based as well as temporal networks.
\\ \ \\
\noindent{\bf DynSI}: this model adapt the classical formulation of SI model (where the transition is $S\rightarrow I$) to the snapshot based topology evolution where the network structure is updated during each iteration.
The model applied at day $t_i$ will then use as starting infected set the result of the iteration performed on the interaction graph of the previous day, and as social structure the current one. 
Such choice implies that not only the interactions of consecutive snapshot could vary but that the node sets can also differ.\\

\noindent{\bf DynSIS}: as the previous dynamic model, the DynSIS adapt the classical formulation of SIS model (where the transition is $S\rightarrow I\rightarrow S$) to the snapshot based topology evolution where the network structure is updated during each iteration.
The DynSIS implementation assumes that the process occurs on a directed/undirected dynamic network.\\

\noindent{\bf DynSIR}: as the previous model, the DynSIS adapt the classical formulation of SIR model (where the transition is $S\rightarrow I\rightarrow R$) to the snapshot based topology evolution where the network structure is updated during each iteration.
The DynSIR implementation assumes that the process occurs on a directed/undirected dynamic network.

\section{Opinion Dynamic Models.}

{\bf Voter}: this model is one of the simplest models of opinion dynamics, originally introduced to analyze competition of species \cite{clifford73} and soon after  applied to model elections \cite{holley75}. 
The model assumes the opinion of an individual to be a discrete variable $\pm 1$.
The state of the population varies based on a very simple update rule: at each iteration, a random individual is selected, who then copies the opinion of one random neighbor. 
Starting from any initial configuration, on a complete network, the entire population converges to a consensus on one of the two options  \cite{krapivsky2010kinetic}. 
The probability that consensus is reached on opinion $+1$ is equal to the initial fraction of individuals holding that opinion.\\

\noindent{\bf Snajzd}: this model \cite{Sznajd-Weron2001} is a variant of spin model employing the theory of social impact, which takes into account the fact that a group of individuals with the same opinion can influence their neighbors more than one single individual. 
In the original model the social network is a 2-dimensional lattice, however, we also implemented the variant on any complex networks. 
Each agent has an opinion $\sigma_i=\pm 1$; at each time step, a pair of neighboring agents is selected and, if their opinion coincides, all their neighbors take that opinion. 
The model has been shown to converge to one of the two agreeing stationary states, depending on the initial density of up-spins (transition at 50\% density). \\

\noindent{\bf Q-Voter}: this model was introduced as a generalization of discrete opinion dynamic models \cite{Castellano2009}.  
Here, N individuals hold an opinion $\pm 1$. 
At each time step, a set of $q$ neighbors are chosen and, if they agree, they influence one neighbor chose at random, i.e. this agent copies the opinion of the group. 
If the group does not agree, the agent flips its opinion with probability $\epsilon$.  
It is clear that the voter and  Sznajd models are special cases of this more recent model ($q=1, \epsilon=0$ and $q=2, \epsilon=0$, respectively). 
Analytic results for $q\leq3$ validate the numerical results obtained for the special case models, with transitions from an ordered phase (small $\epsilon$) to a disordered one (large $\epsilon$). 
For $q>3$, a new type of transition between the two phases appears, which consist of passing through an intermediate regime where the final state depends on the initial condition. 
We implemented in NDlib the model with $\epsilon=0$.  \\

\noindent{\bf Majority Rule}: this model is a different discrete model of opinion dynamics, proposed to describe public debates \cite{galam02}. 
Agents take discrete opinions $\pm 1$, just like the voter model.
All agents can interact with all other agents (also in our implementation), so the social network is always a complete graph. 
At each time step, a group of $r$ agents is selected randomly and they all take the majority opinion within the group. 
The group size can be fixed or taken at each time step from a specific distribution. 
If $r$ is odd, then the majority opinion is always defined, however, if $r$ is even there could be tied situations. 
To select a prevailing opinion, in this case,  a bias in favor of one opinion ($+1$) is introduced. 
This idea is inspired by the concept of social inertia \cite{friedman84}. \\	   

\noindent{\bf Cognitive Opinion Dynamics}: this model was introduced by Vilone et all. \cite{vilone2016reducing}, which models the state of individuals taking into account several cognitively-grounded variables. 
The aim is to simulate a response to risk in catastrophic events in the presence of external (institutional) information. 
The individual opinion is modeled as a continuous variable $O_i \in [0,1]$, representing the degree of perception of the risk (how probable it is that the catastrophic event will actually happen). 
This opinion evolves through interactions with neighbors and external information, based on four internal variables for each individual $i$:  risk sensitivity ($R_i \in \{-1,0,1\}$), tendency to inform others ($\beta_i \in [0,1]$), trust in institutions ($T_i \in [0,1]$) and trust in peers ($\Pi_i = 1-T_i$). 
These values are generated when the population is initialized and stay fixed during the simulation. 
In our implementation, we allow some control on the distribution of these parameters. 
The update rules define how $O_i$ values change in time (see original paper \cite{vilone2016reducing} for details). 
The model was shown to be able to reproduce well various real situations; in particular, it is visible that risk sensitivity is more important than trust in institutional information when it comes to evaluating risky situations.

\bibliographystyle{spmpsci}      
\bibliography{rel}

\clearpage

\end{document}